\documentclass{emulateapj}
\shorttitle{Atmospheric Phase Characteristics of ALMA Long Baseline}
\shortauthors{Matsushita et al.}

\begin{document}

\title{ALMA Long Baseline Campaigns: Phase Characteristics of
	Atmosphere at Long Baselines in the Millimeter and Submillimeter
	Wavelengths}

\author{Satoki Matsushita\altaffilmark{1},
		\email{satoki@asiaa.sinica.edu.tw}
	Yoshiharu Asaki\altaffilmark{2,3},
	Edward B. Fomalont\altaffilmark{2,4},
	Koh-Ichiro Morita\altaffilmark{2,3},
	Denis Barkats\altaffilmark{5},
	Richard E. Hills\altaffilmark{6},
	Ryohei Kawabe\altaffilmark{7,8,9},
	Luke T. Maud\altaffilmark{10},
	Bojan Nikolic\altaffilmark{6},
	Remo P. J. Tilanus\altaffilmark{10},
	Catherine Vlahakis\altaffilmark{4},
	Nicholas D. Whyborn\altaffilmark{2}
	}

\altaffiltext{1}{Academia Sinica Institute of Astronomy and
	Astrophysics, P.O.\ Box 23-141, Taipei 10617, Taiwan, R.O.C.}
\altaffiltext{2}{Joint ALMA Observatory, Alonso de C\'ordova 3107,
	Vitacura 763 0355, Santiago, Chile}
\altaffiltext{3}{Chile Observatory, National Astronomical Observatory
	of Japan, National Institutes of Natural Sciences, Joaquin Montero
	3000 Oficina 702, Vitacura, Santiago, C.P.7630409, Chile}
\altaffiltext{4}{National Radio Astronomy Observatory, 520 Edgemont
	Rd, Charlottesville, VA 22903, USA}
\altaffiltext{5}{Harvard-Smithsonian Center for Astrophysics,
	60 Garden St., MS-78, Cambridge, MA 02138, USA}
\altaffiltext{6}{Astrophysics Group, Cavendish Laboratory, University
	of Cambridge, JJ Thomson Avenue, Cambridge CB3 0HE, UK}
\altaffiltext{7}{National Astronomical Observatory of Japan, 2-21-1
	Osawa, Mitaka, Tokyo 181-8588, Japan}
\altaffiltext{8}{Department of Astronomy, School of Science,
	University of Tokyo, Bunkyo, Tokyo 113-0033, Japan}
\altaffiltext{9}{SOKENDAI (The Graduate University for Advanced
	Studies), 2-21-1 Osawa, Mitaka, Tokyo 181-8588, Japan}
\altaffiltext{10}{Leiden Observatory, Leiden University, P.O.\ Box
	9513, 2300 RA Leiden, The Netherlands}

\begin{abstract}
We present millimeter- and submillimeter-wave phase characteristics
measured between 2012 and 2014 of Atacama Large
Millimeter/submillimeter Array (ALMA) long baseline campaigns.
This paper presents the first detailed investigation of the
characteristics of phase fluctuation and phase correction methods
obtained with baseline lengths up to $\sim15$ km.
The basic phase fluctuation characteristics can be expressed with the
spatial structure function (SSF).
Most of the SSFs show that the phase fluctuation increases as a
function of baseline length, with a power-law slope of $\sim0.6$.
In many cases, we find that the slope becomes shallower (average of
$\sim0.2-0.3$) at baseline lengths longer than $\sim1$ km, namely
showing a turn-over in SSF.
These power law slopes do not change with the amount of precipitable
water vapor (PWV), but the fitted constants have a weak correlation
with PWV, so that the phase fluctuation at a baseline length of 10 km
also increases as a function of PWV.
The phase correction method using water vapor radiometers (WVRs)
works well, especially for the cases where PWV $>1$ mm, which reduces
the degree of phase fluctuations
by a factor of two in many cases.
However, phase fluctuations still remain after the WVR phase
correction, suggesting the existence of other turbulent constituent
that cause the phase fluctuation.
This is supported by occasional SSFs that do not exhibit any
turn-over;
these are only seen
when the PWV is low (i.e., when the WVR phase correction works less
effectively) or after WVR phase correction.
This means that the phase fluctuation caused by this turbulent
constituent is inherently smaller than that caused by water vapor.
Since in these rare cases there is no turn-over in the SSF up to the
maximum baseline length of $\sim15$ km, this turbulent constituent
must have scale height of 10 km or more, and thus cannot be water
vapor, whose scale height is around 1 km.
Based on the characteristics, this large scale height turbulent
constituent is likely to be water ice or a dry component.
Excess path length fluctuation after the WVR phase correction at a
baseline length of 10 km is large ($\gtrsim200~\mu$m), which is
significant for high frequency ($>450$~GHz or $<700~\mu$m)
observations.
These results suggest the need for an additional phase correction
method to reduce the degree of phase fluctuation, such as fast
switching, in addition to the WVR phase correction.
We simulated the fast switching phase correction method using
observations of single quasars, and the result suggests that it works
well, with shorter cycle times linearly improving the coherence.
\end{abstract}


\section{Introduction}
\label{sect-intro}

The Atacama Large Millimeter/submillimeter Array
\citep[ALMA;][]{hil10} is the world's largest
millimeter/submillimeter (mm/submm) interferometer.
One of the most important technical developments for ALMA full
operation has been making the longest ($\sim15$ km) baseline
observations possible, which was achieved for the first time at the
end of 2014 \citep{fom15}.

Prior to that, the longest baseline observations for mm/submm linked
arrays had been much shorter, only $\sim2$ km at 230 GHz with the
Berkeley-Illinois-Maryland Association (BIMA) or the Combined Array
for Research in Millimeter-wave Astronomy (CARMA), and several
hundred meters at 345 GHz or higher frequencies with the
Submillimeter Array (SMA) or the NOrthern Extended Millimeter Array
(NOEMA), although there are some very long baseline interferometry
(VLBI) observations with the frequency up to 230 GHz and with
baseline lengths up to $\sim4000$ km \citep[e.g.,][]{doe08,doe12}.
The longest baseline length currently possible with ALMA ($\sim15$
km), together with a significant increase of the number of baselines
(1225 baselines with 50 antennas) compared with the aforementioned
mm/submm arrays, is therefore a revolutionary improvement for
mm/submm linked arrays.
This 15 km baseline length with good $uv$ coverage provides
milli-arcsecond resolution; such high sensitivity and high fidelity
at these wavelengths cannot be obtained with any other current
facilities.
Indeed, in the ALMA long baseline campaign in 2014, science
verification images revealed extraordinally detailed features of an
asteroid \citep{hun15}, a protoplanetary disk system \citep{bro15},
and a gravitationally lensed system \citep{vla15}.
In order to continue to improve on the quality of the data, it is
important to characterize the atmospheric phase fluctuation, and to
test phase correction methods at these long baselines.

Characterization of the atmospheric phase fluctuation so far has
mainly been carried out at low frequencies; Very Large Array (VLA)
was often used to characterize the atmospheric phase fluctuation at
centimeter-wave (cm-wave) with baseline lengths up to a few tens
of km \citep{sra90,car99}.
At higher frequencies, studies with $80-230$ GHz ($1-3$ mm) are
available with baseline lengths up to only about 1 km
\citep{wri96,asa98,mat10}.
These previous cm-/mm-wave studies displayed similar results for
cases where the atmospheric phase fluctuation is caused by water
vapor in the atmosphere.

Since 2010, as a part of ALMA Commissioning and Science Verification
(CSV), and most recently, as part of the Extension and Optimization
of Capabilities \citep[EOC; see][]{fom15}, we have conducted several
ALMA long baseline campaigns, starting from the longest baseline
length of 600 m in 2010 and 2011, 2 km in 2012, 3 km in 2013, and
finally $10-15$ km in 2014.
In the earlier long baseline campaigns, we mainly conducted basic
tests, such as characterizing the phase fluctuation and checking the
effectiveness of the WVR phase correction; in the later phase, we
mainly concentrated on the coherence time calculation and the
evaluation of the fast swtiching phase correction method.
Some of the early test results have been reported in
\citet{mat12,mat14,mat16} and \citet{asa12,asa14,asa16}, and the
overview of the latest $10-15$ km baseline test has been reported in
\citet{fom15}.

In this paper, we present the detailed characterization of phase
fluctuation, improvement of phase fluctuation after the water vapor
radiometer (WVR) phase correction method \citep{wie01,nik13},
coherence time calculation, and the cycle time analysis for the fast
switching phase correction method using the data obtained in the ALMA
long baseline campaigns.
The results presented in this paper replace those reported previously
by \citet{mat12,mat14,mat16}.
Note that the relation between phase fluctuation and the weather
parameters (wind speed, wind direction, temperature, pressure, etc.)
will be discussed in the forthcoming paper (Maud et al., in prep.),
so that we do not discuss here.

\section{Observations and Data Reduction}
\label{sect-obs}


In this paper, we use the data taken in the long baseline campaigns
between 2012 and 2014 (see Sect.~\ref{sect-app-data}).
All the data were taken with observations of a strong point source
(i.e., radio-loud quasars) for tens of minutes (usually $10-40$
minutes, depending on the longest baseline length) with $\sim1$ s
integration time per data point.
Hereafter we refer to these observations as ``single source stares.''
For each measurement, ten or more antennas with various baseline
lengths have been used.
We only used 12 m diameter antenna data and did not use data taken
with 7 m diameter antennas in order to make the thermal noise
contribution uniform over the data as much as possible.

The single source stares were taken for the purpose of statistical
phase analysis beyond the wind crossing time of the longest baseline.
For example, assuming
(1) a longest baseline length of 10 km,
(2) a wind speed along a given baseline of 10 m s$^{-1}$, and
(3) that the phase screen does not change with time
\citep[i.e., frozen phase screen;][]{tay38,dra79},
then the crossing time can be calculated as 1000 s ($\sim17$ minutes).
To have statistically significant data for the longest baseline data,
about twice the measurement time is needed (see \citealt{asa96} for
the detailed discussion)\footnotemark[11].
\footnotetext[11]{Longer measurement time of 90 minutes has also been
tested; see Sect.~\ref{sect-ssf-slope}.}
We used Bands 3 (100 GHz band), 6 (200 GHz band), and 7 (300 GHz
band) for all the campaigns, except for Band 8 (400 GHz band), which
was only used in the 2014 campaign.

The data were reduced using the Common Astronomy Software
Applications (CASA) package \citep{mcm07}, using a standard ALMA
calibration prescription.
Below we give a brief overview.
The WVR phase correction\footnotemark[12]
\footnotetext[12]{WVR phase correction method is to estimate the
amount of water vapor in front of each antenna using the 183 GHz WVR
\citep{nik13}, calculate the excess path length and phase difference
between antennas, then correct the phase fluctuation caused by water
vapor \citep{wie01,nik13}.}
was applied using the program {\it wvrgcal}
\citep{nik12} installed inside CASA.
The precipitable water vapor (PWV) values calculated by {\it wvrgcal}
is used as a measure of the water vapor content in front of the array
of each data set.

Linear phase drift, which is mostly caused by the antenna position
determination error
(main reason for this error is due to poor knowledge of the dry air
 delay term for each antenna, which can be very different from each
 other due to the height difference; see \citealt{fom15} for details),
has been removed from each dataset (the antenna
position errors are generally small enough that they only cause
linear phase drift on timescales of tens of minutes or more).
After this, we averaged the data points for 10 s to take out phase
errors due to thermal noise.
With this 10 s time averaging, and assume the source flux density is
1 Jy and the system temperature is a typical value for each ALMA Band
\citep{rem15}, root mean square (rms) phase errors due to thermal
noise are calculated as 3.4, 2.0, 2.4, and $4.5~\mu$m for Bands 3, 6,
7, and 8, respectively.
In our observations, the minimum source flux densities are 1.7, 1.1,
1.2, and 3.7 Jy for Bands 3, 6, 7, and 8, respectively, so that the
maximum rms phase errors can be calculated as 2.0, 1.8, 2.0, and
$1.2~\mu$m for Bands 3, 6, 7, and 8, respectively.
The smallest measured phase fluctuation at very short baseline
lengths of about 15 m before the WVR phase correction, but after the
10 s averaging, is about 6 $\mu$m \citep{mat12}, which is larger than
the upper limit of the phase error.
We can therefore conclude that we successfully suppressed the effect
of instrumental noise in our data with 10 s averaging.

We then calculated the rms phase fluctuation for each baseline.
For the unit of phase, we express in path length\footnotemark[13].
\footnotetext[13]{
$\Phi = \frac{\theta}{360\arcdeg} \times \frac{c}{\nu}$, where $\Phi$
and $\theta$ are phase in path length and in degrees, respectively,
$c$ is the speed of light, and $\nu$ is the observation frequency.}
This allows us to directly compare the results from various
frequencies in the same unit without any frequency dependence, and is
often referred to as excess path length.

\section{Results and Discussion}
\label{sect-res}

\subsection{Improvement Factor of the WVR Phase Correction}
\label{sect-wvr}


\begin{figure}
\plotone{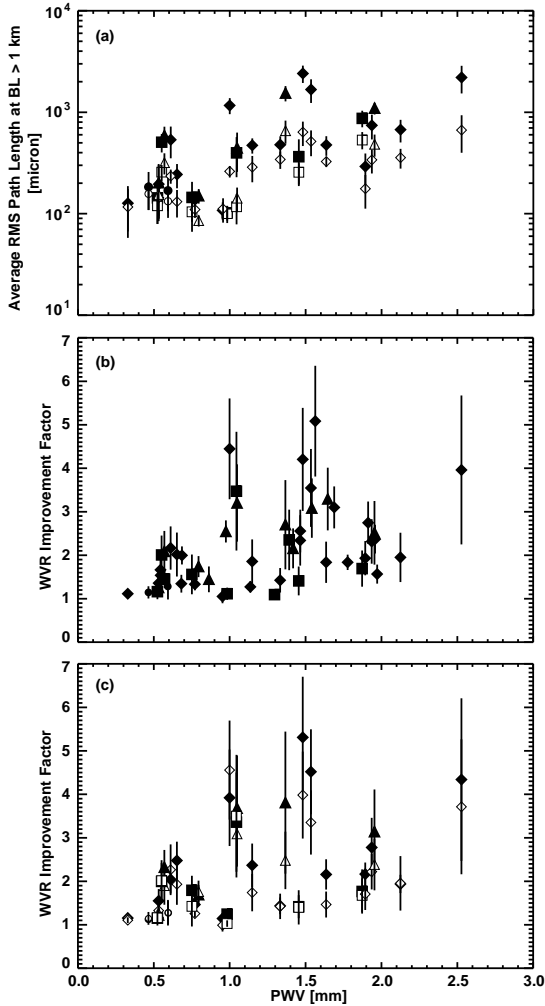}
\caption{(a) Averaged rms excess path length at baseline lengths
	longer than 1 km.
	This plot only uses the 2014 data, since only the 2014 data have
	statistically significant data points at baselines longer than
	1 km.
	Symbols are differentiated with the frequency bands; diamond,
	triangle, square, and circle symbols indicate the data taken with
	Bands 3, 6, 7, and 8, respectively.
	Filled and open symbols are before and after the WVR phase
	correction, respectively.
	(b) Improvement factor of the WVR phase correction (ratio of the
	rms excess path length without WVR phase correction to that with
	the correction) as a function of PWV.
	Each data point is averaged over all baselines.
	This plot uses all the $2012-2014$ data.
	(c) Data points have been separated for the one averaged over the
	baseline length shorter (filled) or longer (open) than 1 km.
	The open symbol data points are calculated from the data shown in
	(a).
\label{fig-wvr}}
\end{figure}

WVR phase correction often improves the data quality
\citep{mat12,mat14,nik13,fom15}, but there is no statistical study
how much it improves for short to long baselines.
In this subsection, we present the statistical study results for this
method.

Fig.~\ref{fig-wvr}(a) shows the rms excess path length before (filled
symbols) and after (open symbols) the WVR phase correction for the
data with baseline length longer than 1 km (data for baseline length
$>1$ km in eahc data set have been averaged into one data point).
It is obvious that the WVR phase correction reduces all the rms
excess path lengths to less than $1000~\mu$m.

The improvement factor of the WVR phase correction method has been
calculated as the ratio of the rms excess path length for each
baseline without the WVR phase correction to that with the
correction.
Higher values mean that the WVR phase correction works better
(improved more) than cases with lower values (less improvement).
Values of less than unity means that the WVR phase correction made
the phase fluctuation worse than the original data.
Calculated values are averaged over all baselines, and plotted as a
function of PWV (Fig.~\ref{fig-wvr}b).
In addition, we plotted the data divided into baseline lengths
shorter and longer than 1 km (Fig.~\ref{fig-wvr}c).
For the former, we used data from all campaigns; for the latter, we
use only data from the 2014 campaign, since the older data contain
very few long baselines.

The mean improvement factor over all PWV conditions is $2.1\pm0.7$.
This result is consistent with what is reported previously
\citep{mat12,fom15}, but this is the first statistical result.
However, from Fig.~\ref{fig-wvr}(b), it is obvious that the
improvement factor is often larger when PWV is larger than 1 mm, with
a mean improvement factor of $2.4\pm0.7$.
In the case of PWV $<$ 1 mm, however, the ratio is $1.7\pm0.4$.
Thus it is clear that although the WVR phase correction in most cases
provides an improvement in the amount of phase fluctuation, the
amount of improvement can be significantly larger for PWV $>$ 1 mm.
This result can be explained because at low PWV, the difference in
the amount of water vapor along the lines of sight of two antennas is
small compared to the thermal noise of the WVRs of $\sim0.1$ K rms
\citep[which corresponds to about a few tens micron rms;][]{nik13}.
In addition, scatter of the data points is also different between low
and high PWV cases, which is obvious in Fig.~\ref{fig-wvr}(b) and
also from the error values of about twice difference ($\pm0.7$ vs
$\pm0.4$) as mentioned above.
This means that not all the data show significant improvement in
phase fluctuation in the case of PWV $>$ 1 mm after the WVR phase
correction.

Fig.~\ref{fig-wvr}(c) shows that
the improvement factor for the baseline length shorter than 1 km is
$2.3\pm0.9$, and that for the baseline length longer than 1 km is
$2.0\pm0.7$.
The longer baseline data tend to have lower improvement factor than
the shorter ones, although the difference is statistically not
significant.

In both plots, different frequency bands are plotted with different
symbols, but we find no significant difference in the improvement
factor as a function of frequency band.

\vspace{3ex}

\subsection{Spatial Structure Function of Phase Fluctuation}
\label{sect-ssf}

In this subsection, we first define the spatial structure function
(SSF) of phase (Sect.~\ref{sect-ssf-def}), and derive the SSF slopes
(Sect.~\ref{sect-ssf-slope}) and constants
(Sect.~\ref{sect-ssf-phase}) for all the past long baseline single
source stare data.
Statistical result for the slopes is compared with the previously
published studies (Sect.~\ref{sect-ssf-slope}).
We then estimate the rms phase fluctuation for a 10 km baseline, and
derive a weak correlation with PWV (Sect.~\ref{sect-ssf-phase}).
After WVR phase correction, there is a significant residual phase
fluctuation,
suggesting that there may be other constituents than water vapor
that cause the phase fluctuation exist in the atmosphere; we discuss
the possibilities for liquid water, ice, and dry components, and also
any instrumental causes (Sect.~\ref{sect-ssf-why}).

\subsubsection{Definition of Spatial Structure Function}
\label{sect-ssf-def}

The SSF of phase is defined as
\begin{equation}
D_{\theta}(d) = <\{\theta(x)-\theta(x-d)\}^{2}>,
\label{eq-ssf}
\end{equation}
where $\theta(x)$ and $\theta(x-d)$ are phases at positions $x$ and
$x-d$, and the angle brackets mean an ensemble average
\citep{tat61,tho01}.
Since $\theta(x)-\theta(x-d)$ is an interferometer phase with
a baseline length $d$,
$\sqrt{D_{\theta}(d)}$ is approximated to the rms phase fluctuation
of a baseline over the entire observation time.
SSF plots in this paper have therefore been made by calculating
the rms phase
for each projected baseline in the direction of a target source
using tens of minutes of observation time, and plotted as a function
of baseline length.

Slopes for rms phase fluctuation are well studied theoretically, and
in the case of 3-dimensional (3-D) Kolmogolov turbulence of the
Earth's atmosphere, it is expected to have a slope of 0.83;
for the 2-D turbulence case, the slope is expected to be 0.33.
In the case of no correlation in the atmospheric turbulence between
two antennas, the slope is expected to be zero \citep{tho01}.

\begin{figure}
\plotone{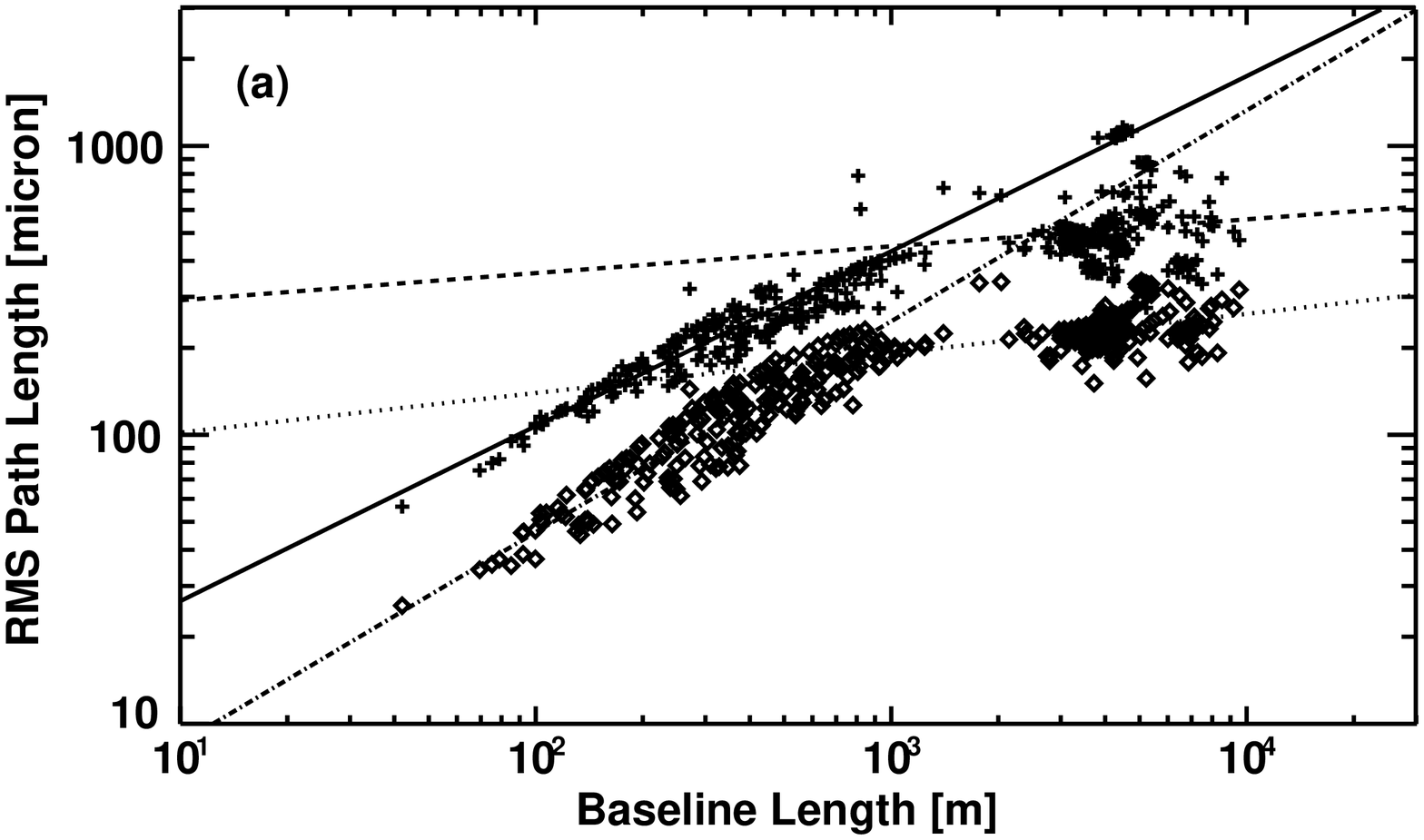}
\plotone{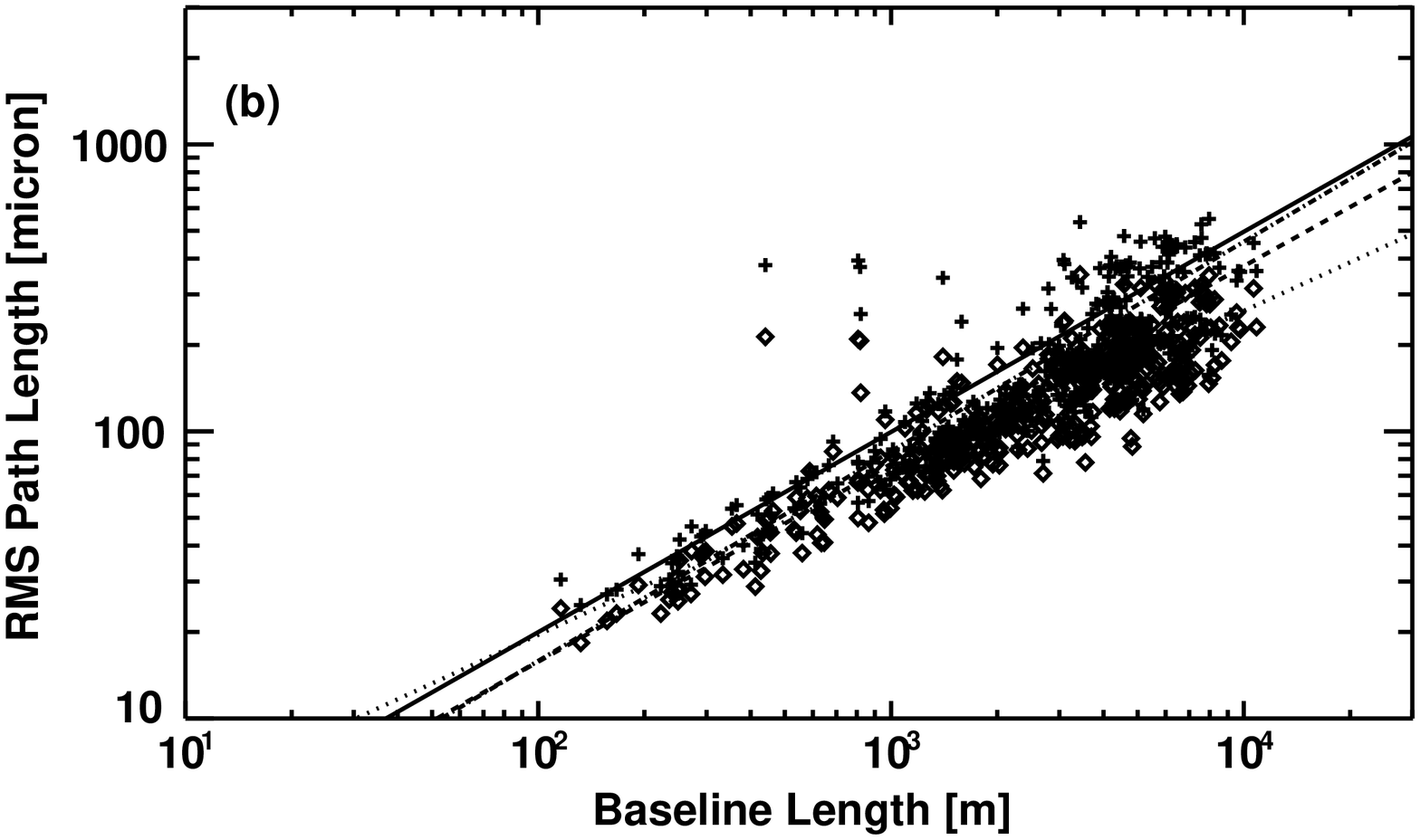}
\caption{Two representative plots of the spatial structure function
	(SSF) with (open diamond) and without (cross) WVR phase
	correction.
	Solid and dashed lines are the fitting results of slopes using
	baselines shorter than 500 m and longer than 1 km, respectively.
	(a) Example of SSF with a clear turn-over.
	The data file for this plot is ``uid\_\_\_A002\_X8bd3e8\_Xa59''
	in Table~\ref{tab-app-10km}.
	(b) Example of SSF without any clear turn-over.
	The data file for this plot is ``uid\_\_\_A002\_X8e1004\_Xf04''
	in Table~\ref{tab-app-10km}.
\label{fig-ssf}}
\end{figure}

Fig.~\ref{fig-ssf} shows two example plots of SSFs we obtained using
the ALMA data.
In these plots, both the WVR phase corrected and uncorrected SSFs are
displayed.
As mentioned above, phase is converted into the unit of path length.
Fig.~\ref{fig-ssf}(a) displays a typical SSF plot; there is a clear
turn-over in the SSF at baseline lengths of $\sim1$ km in both plots
with and without the WVR phase correction.
Shorter baselines, before the turn-over, show a steeper slope, while
longer baselines show a shallow or almost flat slope.
In general, for the `typical' data sets, only one turn-over is
evident with the current data analysis method, and is usually
observed between baseline lengths of several hundred meters to
$\sim1$ km.
Fig.~\ref{fig-ssf}(b), on the other hand, exhibits no clear
turn-over; phase fluctuation increases constantly even at long
baselines.
Such a constant slope SSF is not common, but is sometimes observed in
the data (see Sect.~\ref{sect-ssf-slope} for statistics).
Note that these no turn-over data generally have a much lower excess
path length value compared to the `typical' data at particular
baseline lengths (especially at shorter baselines), which can be seen
in Fig.~\ref{fig-ssf}.
This means that the SSF with a constant slope is inherently more
stable than that with a turn-over (see also the end of
Sect.~\ref{sect-ssf-slope}).

The baseline length of the turn-over roughly corresponds to the
scale height of the three-dimensional atmospheric turbulence
\citep{tre87,lay97}.
Our results suggest that the scale height of the water vapor
constituent
at the ALMA site is about 1 km in most cases, which is a
typical value of the turn-over baseline length measured with
astronomical radio arrays \citep[e.g.,][]{car99}.
It is highly possible that the main turbulent constituent at the ALMA
site is water vapor,
which is consistent with the improvement of
phase fluctuation by the WVR phase correction in most cases
(Sect.~\ref{sect-wvr}).
On the other hand, no turn-over means that the scale height of the
turbulent constituent is higher than 10 km.
This scale height is much higher than that of water vapor, suggesting
that the turbulent constituent for the SSF with no turn-over is
caused by some other constituent (see Sect.~\ref{sect-ssf-why} for
further discussion).

\subsubsection{Slopes for the Spatial Structure Functions}
\label{sect-ssf-slope}

\begin{figure}
\plotone{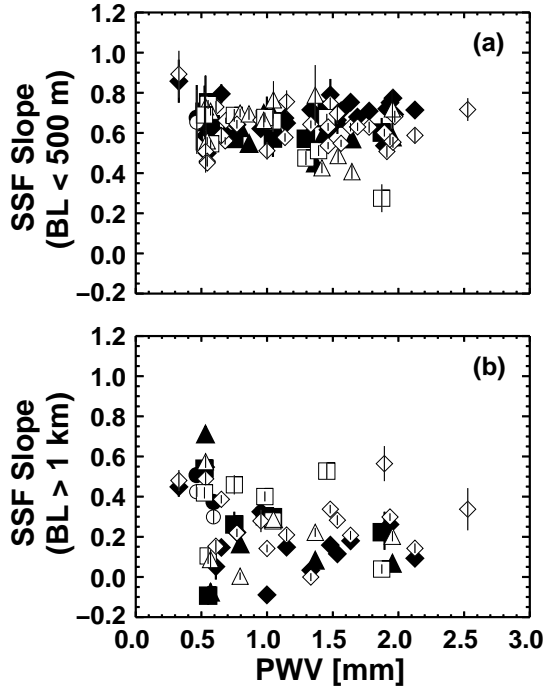}
\caption{Fitted slopes of the spatial structure functions (SSFs) as a
	function of PWV.
	(a) Slopes for the data points in each SSF with the baseline
	lengths shorter than 500 m, and (b) longer than 1 km.
	Symbols are differentiated with the frequency bands; diamond,
	triangle, square, and circle symbols indicate the data taken with
	Bands 3, 6, 7, and 8, respectively.
	Filled symbols are the data before the WVR phase correction, and
	open symbols after.
	Note that (a) includes the $2012-2014$ data, but (b) only
	includes the 2014 data, for the reason same as in
	Sect.~\ref{sect-wvr}.
\label{fig-slope}}
\end{figure}

\begin{deluxetable*}{ccccccc}
\tablecaption{Fitted slopes of the spatial structure functions
	(SSFs).
	\label{tab-ssf-slope}}
\tablehead{
	\colhead{}
		& \multicolumn{3}{c}{{Without WVR phase correction}}
		& \multicolumn{3}{c}{{With WVR phase correction}} \\
	\colhead{Baseline Length}
		& \colhead{All}
		& \colhead{PWV $<1$ mm}
		& \colhead{PWV $>1$ mm}
		& \colhead{All}
		& \colhead{PWV $<1$ mm}
		& \colhead{PWV $>1$ mm}
	}
\startdata
$<500$ m  & $0.65\pm0.06$ & $0.65\pm0.06$ & $0.66\pm0.06$
          & $0.62\pm0.09$ & $0.64\pm0.08$ & $0.60\pm0.09$ \\
  $>1$ km & $0.22\pm0.15$ & $0.27\pm0.20$ & $0.17\pm0.07$
          & $0.29\pm0.13$ & $0.31\pm0.15$ & $0.26\pm0.11$
\enddata
\end{deluxetable*}

We first fitted the slopes for the data points with baseline lengths
shorter than 500 m and longer than 1 km for each SSF plot, namely
shorter and longer than the turn-overs in the plots.
The fitting function is expressed as
\begin{equation}
\log_{10}(\Delta L) = a \times \log_{10}(d) + b,
\label{eq-path-fitting}
\end{equation}
where $\Delta L$ is the rms excess path length in micron, $d$ is the
baseline length in meter, and $a$ and $b$ are the slope, namely the
structure exponent, and the structure constant, respectively.
Examples of the fitting are shown in Fig.~\ref{fig-ssf}, and the
fitted slopes as a function of PWV are plotted in
Fig.~\ref{fig-slope}; (a) is for the shorter baseline slopes, and
(b) for the longer baseline slopes.

In Fig.~\ref{fig-slope}(a), there is no significant trend either
as a function of PWV, frequency, or the effect of the WVR phase
correction.
The fitted slopes before and after the WVR phase correction with
averaging all the data, and that with PWV lower and higher than 1 mm
are shown in the top row of Table~\ref{tab-ssf-slope}.
All those values are around 0.6, indicating that the WVR phase
correction or the amount of water vapor in the atmosphere does not
change the shorter baseline slopes of SSFs.
The average slope without the WVR phase correction of $0.65\pm0.06$
is consistent with the 50\% quartile slope for the 3-year (1996 July
-- 1999 March) statistical data using the site testing 11.2 GHz radio
seeing monitor of 0.63 \citep{but01}.
This indicates that the data we took are typical phase fluctuation
characteristics at the ALMA site.

Fig.~\ref{fig-slope}(b) clearly shows that the slopes at longer
baselines (baseline length $> 1$ km) are shallower than the shorter
baselines (baseline length $< 500$ m; Fig.~\ref{fig-slope}a) in most
cases.
Similar to the above, the fitted slopes are shown in the bottom row
of Table~\ref{tab-ssf-slope}.
The average slopes are $0.22\pm0.15$ and $0.29\pm0.13$ for the data
before and after the WVR phase correction, respectively.
These values are significantly smaller than those for the baseline
length shorter than 500 m, indicating that most of the data have a
turn-over at a baseline length between 500 m and 1 km.

For almost all the cases, the slopes for shorter baselines are in the
middle of theoretical 3-D and 2-D Kolmogorov turbulence, which are
0.83 and 0.33, respectively.
This suggests the existence of undeveloped turbulence or multi-layer
turbulence with different heights in the atmosphere.
The slopes for longer baselines (slopes after the turn-over), on the
other hand, exhibit the values between the 2-D Kolmogorov turbulence
(slope = 0.33) and no correlation between two antennas (slope = 0) in
most cases.
%

Our result is very similar to the result from the statistical study
of the SSF slope at the VLA site; the slope of 0.59 at short
($<1$ km) baselines, and 0.3 at longer baselines \citep{sra90}.
On the other hand, later SSF study with long measurement time of
90 minutes \citep{car99} exhibits two clear turn-overs, one from the
3-D Kolmogorov turbulence (slope = 0.83) to the 2-D one (slope =
0.33), and the other from the 2-D one to the no-correlation regime
(slope = 0).
SSFs with two turn-overs has never been obtained in the statistical
study by \citet{sra90}, suggesting that SSFs with two turn-overs are
statistically rare, the coverage of the array configuration was not
wide enough to investigate the full range of the atmospheric
turbulence, or the measurement time was not long enough.
Note that we took one 90 minutes long data (see
``uid\_\_\_A002\_X8e1004\_Xfe5'' in Table~\ref{tab-app-10km}), but
this data set did not show two turn-overs as \citet{car99}, but
showed similar feature as the 30 minutes long data taken in the same
day.


For PWV $< 1$ mm, the average of the slope at longer baselines is
much smaller than that for the shorter baselines, but the scatter of
the slope is large (Table~\ref{tab-ssf-slope}), and the highest
values of the slope are almost the same as that of the shorter
baselines.
This means that 
there are some cases that do not have any turn-over; an example is
shown in Fig.~\ref{fig-ssf}(b).
As mentioned in Sect.~\ref{sect-wvr}, at this low PWV range, the
improvement factor of the WVR phase correction is not high,
suggesting that the phase fluctuation is not due to water vapor but
caused by another constituent, the same conclusion as above
(Sect.~\ref{sect-ssf-def}).

For PWV $> 1$ mm and without the WVR phase correction, the slopes are
always small (small values with small scatter;
Table~\ref{tab-ssf-slope}), but after the WVR phase correction, both
the slope and the scatter are large.
This, together with the improvement factor of the WVR phase
correction mentioned above (Sect.~\ref{sect-wvr}), indicates that in
many of the cases the slopes turn to be steeper after the WVR phase
correction for the longer ($> 1$ km) baselines.
For some cases, the slopes become similar to the slope at shorter
($< 500$ m) baselines, namely SSF being similar to the one with no
turn-over, suggesting that the WVR phase correction took out all the
phase fluctuation caused by water vapor, and that caused by another
constituent remains.
This means that if we can completely remove the phase fluctuation due
to water vapor in the atmosphere, a large-scale turbulent component
that covers the longest baseline lengths is revealed.
This also means that phase fluctuation due to the large-scale
turbulent component is inherently smaller than that due to water
vapor.

\subsubsection{Constants for the Spatial Structure Functions}
\label{sect-ssf-const}

\begin{figure}
\plotone{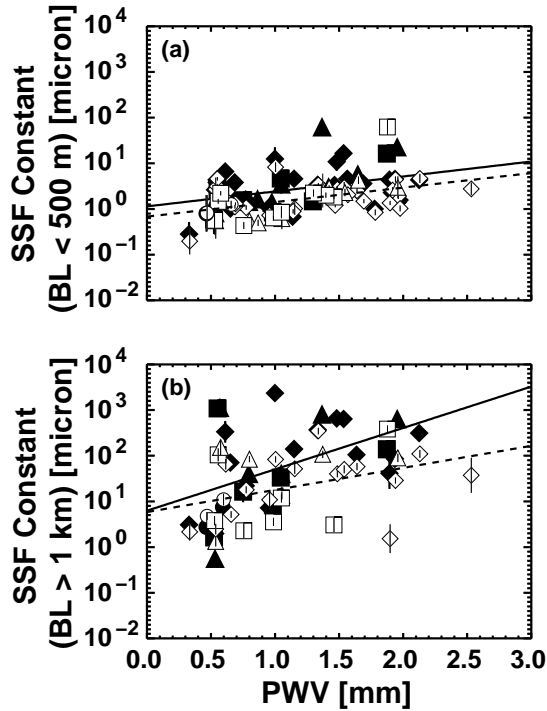}
\caption{Fitted structure constants of SSFs as a function of PWV.
	Symbols and their colors are the same as in Fig.~\ref{fig-slope}.
	(a) Structure constants for the data points in each SSF with the
	baseline lengths shorter than 500 m.
	Solid [$b_{raw} = (0.3\pm0.1) \times {\rm PWV} + (0.1\pm0.1)$]
	and dashed
	[$b_{wvr} = (0.3\pm0.1) \times {\rm PWV} - (0.2\pm0.1)$] lines
	are the fitting results for the data before and after the WVR
	phase correction, respectively.
	(b) Structure constants for the data points in each SSF with the
	baseline lengths longer than 1 km.
	Solid [$b_{raw} = (0.9\pm0.3) \times {\rm PWV} + (0.8\pm0.4)$]
	and dashed
	[$b_{wvr} = (0.5\pm0.2) \times {\rm PWV} + (0.8\pm0.3)$] lines
	are the fitting results for the data before and after the WVR
	phase correction, respectively.
\label{fig-const}}
\end{figure}

Figs.~\ref{fig-const}(a) and (b) display the derived structure
constants $b$ in Eq.~\ref{eq-path-fitting} as a function of PWV for
baseline lengths shorter than 500 m and longer than 1 km,
respectively.
For both cases, there is a weak correlation between the structure
constants and PWV for both before and after the WVR phase correction
(the solid and dashed lines, respectively in Fig.~\ref{fig-const}).
With the average slope $a$ in Table~\ref{tab-ssf-slope} and the
fitted results of the structure constant $b$ in Fig.~\ref{fig-const},
it is possible to estimate the rms excess path length on any baseline
length for the current ALMA (i.e., baseline length up to 15 km).
Note that the scatter is large so that this estimation is useful only
to tell the tendency of the phase fluctuation.

\subsubsection{Phase Fluctuation on 10 km Baselines}
\label{sect-ssf-phase}

To understand quantitatively how much phase fluctuation exists at
long baselines, we calculate the rms excess path length for the
baseline length at 10 km.
Although the averaged slope at baseline lengths longer than 1 km is
around $0.2-0.3$ (see Sect.~\ref{sect-ssf-slope}), the slope at
baseline lengths longer than 10 km is expected to be close to zero,
namely no increase of the rms excess path length \citep{car99}.
Therefore it is expected that the rms excess path length at 10 km is
roughly equal to the rms excess path length longer than 10 km.
Here, we estimate the rms excess path length at 10 km using the
fitting results derived in the previous subsection.
This information will be useful for high frequency observations, or
future extended-ALMA \citep{kam13}, which is to extend the maximum
baseline lengths longer than the current ALMA ($>16$ km), or Very
Long Baseline Interferometry (VLBI) observations, which will have
much longer baseline lengths ($>100$ km).

\begin{figure}
\plotone{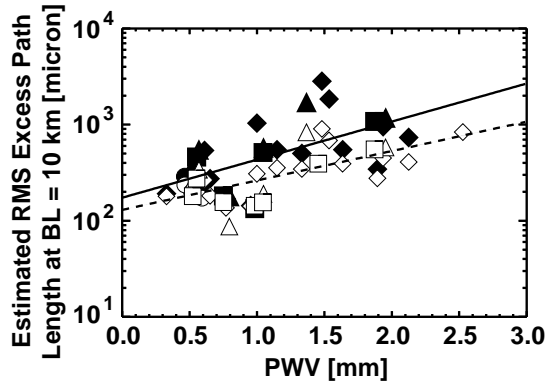}
\caption{Estimated rms excess path length of each data set taken in
	2014 at a baseline length of 10 km using the fitted results
	mentioned in Sect.~\ref{sect-ssf-slope}, which roughly represents
	the rms excess path length even longer than 10 km.
	Solid and dashed lines are the fitting results for the estimated
	rms excess path length before and after the WVR phase correction,
	respectively.
	Symbols and their colors are the same as in Fig.~\ref{fig-slope}.
\label{fig-rms}}
\end{figure}

Using Eq.~(\ref{eq-path-fitting}) to fit the data with baseline
lengths longer than 1 km, we estimated the rms excess path length at
a baseline length of 10 km.
Fig.~\ref{fig-rms} displays the estimated rms excess path length at
a baseline length of 10 km as a function of PWV for both before and
after the WVR phase correction.
As expected from a weak correlation between the structure constant
$b$ and PWV,
both data set
show a weak correlation with PWV (see the solid and dashed lines for
the fitting results); lower PWV conditions tend to have smaller rms
excess path length, and opposite for higher PWV conditions.
This is understandable since a larger amount of water vapor in the
atmosphere provides a greater possibility for larger phase
fluctuations caused by increased differences in the water vapor
content over the array.
On the other hand, the scatter is large, almost an order of magnitude
at a given PWV ranges, which makes the correlation weak.
This indicates that the phase fluctuation cannot only be the fucntion
of the total water vapor content in the atmosphere.

For almost all the data, there is a significant rms excess path
length after the WVR phase correction, with a mean value of
$206~\mu$m even at PWV $<$ 1 mm.
This result indicates that even after the WVR phase correction under
good weather conditions, peak-to-peak phase fluctuation reaches
around $2\pi$ or more at the high frequency bands (i.e., Bands 9 \&
10; $600-1000$ GHz or $300-500~\mu$m) over $\sim10$ minute
timescales.
This phase fluctuation blurs the final synthesized image, with the
blurring size around twice the synthesized beam size or more,
significantly affects the imaging quality, if we calibrate the phase
at the same frequency.
One way to mitigate the $2\pi$ ambiguity problem in the phase
calibration is to use a calibrator phase at lower frequency and to
apply the solutions to the target at higher frequency, i.e., 
band-to-band phase calibration and/or using the fast switching phase
correction method.
The former phase calibration technique has another advantage in ALMA;
because the array sensitivity is higher at lower frequencies, fainter
phase calibrators are available at lower frequency even though the
thermal noise in phase is scaled up with the frequency ratio.
This leads another advantage of the availability of the phase
calibrator; it becomes easier to find a closer phase calibrator to a
science target source.
The fast switching phase correction is further discussed in
Sect.~\ref{sect-fss}.

\subsubsection{Why Does the WVR Phase Correction Not Take Out All the
	Phase Fluctuation?}
\label{sect-ssf-why}

In most cases, the main difference between SSFs with and without the
WVR phase correction is the reduction in the absolute rms phase (rms
path length) values (Fig.~\ref{fig-wvr}a), whereas the overall shapes
and slopes do not change significantly with the WVR phase correction
(Fig.~\ref{fig-ssf}a) and are only weakly correlated with PWV
(Fig.~\ref{fig-rms}).
The WVR phase correction improves the rms path lengths by a factor of
2 on average (Sect.~\ref{sect-wvr}).
If the WVR phase correction takes out all the phase fluctuation, the
resultant spatial structure function should exhibit an almost flat
feature; since the WVR phase correction is applied to the data every
1 second, if the wind blows along a baseline with a velocity of
10 m s$^{-1}$ and assuming a frozen phase screen, then the phase at
the baselines longer than 10 m should show random phase that
corresponds to the thermal noise of the WVRs.
We have never observed such a SSF in the last 5 years of the long
baseline campaigns \citep[see also][]{mat12,mat14,mat16}.
The overall similarity between the SSFs with and without the WVR
phase correction suggests that the origin of the phase fluctuation
is not only due to the water vapor but also other turbulent
constituents.
Furthermore, SSFs without any turn-over also suggests the existence
of other turbulent constituents with scale height larger than the
longest baseline length of $\sim10$ km, as mentioned above
(Sect.~\ref{sect-ssf-def}, \ref{sect-ssf-slope}).

The cause of this remaining phase fluctuation is still unclear:
Turbulent eddies excited by wind blowing over the mountains can cause
additional phase fluctuation.
Indeed, some raw data show this effect, especially data from antennas
near mountains, but the phase fluctuation caused by this effect can
be cleanly removed by the WVR phase correction \citep{asa16}.
Therefore this effect cannot be the cause of the remaining phase
fluctuation.
The remaining phase fluctuation can also be due to wind speed or its
direction parallel or perpendicular to baseline directions.
This will be discussed in the forthcoming paper (Maud et al., in
prep.), so that we do not discuss here.
Liquid water (fog or clouds) in the atmosphere absorbs the amplitude
of electromagnetic waves significantly \citep{ray72,lie89,lie91} at
continuum level.
This changes the line profile of the water vapor in the atmosphere,
so that the WVR phase correction, which uses the liquid water-free
line profile model, will no longer be appropriate
\citep{mat00,mat03}.
Therefore if liquid water dominates the atmosphere, then the rms
phase after WVR phase correction will not improve or will even be
worse.
Since we are looking for the cause of the remaining phase fluctuation
{\it after} successful WVR phase correction, liquid water cannot be
the cause.
Water ice in the atmosphere, which is not detected by the WVR, can
induce similar phase fluctuations to water vapor \citep{huf91,lie93}.
Density fluctuations of a dry component (i.e., N$_{2}$ and O$_{2}$)
in the atmosphere could also be the cause of phase fluctuations
\citep{nik13}.
Both water ice and a dry component have a high scale height of around
10 km or more, so that the SSFs with no turn-over can be explained by
these constituents.
Although it is not clear whether the water ice always exists in the
atmosphere, and the density fluctuations are still not confirmed
observationally, those two atmospheric components could potentially
be the causes.

Instrumentation problems could also be a possibility; since the SSF
is baseline-based, instruments that could cause baseline-based
problems needed to be considered.
The correlator is one of the baseline-base instruments, but it is very
unlikely that this could produce larger phase fluctuation selectively
at longer baselines all the time, because the correlation process
after analog-to-digital conversion of the received signals at each
antenna is highly digitized.
The Line Length Corrector (LLC; round-trip phase corrector) is an
instrument that can create phase noise as a function of baseline
length; noise in this instrumental component causes antenna-based
phase noise, but the atmospheric phase fluctuation is baseline-based,
so that the relative effect of LLC phase noise is larger for shorter
baselines and smaller for longer baselines, namely it appears as
baseline-based.
However, the variation of the LLC in each antenna is always
monitored, and the timescale of the variation is much longer than the
observed phase fluctuation.
This indicates that the LLCs are unlikely to cause the phase
fluctuation.

\subsection{Coherence Time}
\label{sect-coh}

The temporal coherence function is defined as
\begin{equation}
C(T) = \left|\frac{1}{T}\int^{T}_{0}\exp\{-i\theta(t)\}dt\right|,
\end{equation}
where $\theta(t)$ is phase at a time $t$ and $T$ is an arbitrary
integration time \citep{tho01}.
This equation is namely a vector averaging of phase in the time
domain.
A coherence of unity means no loss in coherence (i.e., amplitude) due
to the phase fluctuation, and that of zero means no coherence at all
due to huge phase fluctuation.
Coherence time is a maximum $T$ at which the coherence is not
smaller than a certain critical value;
since phase fluctuation increases as time passes from a
certain time, coherence decreases.

Coherence time is closely connected with an integration time for one
data point for radio arrays; if one can tolerate a certain coherence
loss, one could define the integration time based on the coherence
time.
This is also true for the calibration time interval; for a typical
interferometric observation, a phase calibrator will be observed at
certain time intervals, and if one can tolerate a certain coherence
loss, one could define the calibration time interval based on the
coherence time.

In addition, this is also relevant for the integration time of VLBI;
for detecting interferometric signals (i.e., fringes) with VLBI, it
is important to have high signal-to-noise ratio within one data
point, since it is needed to search the signal in the ranges of delay
and delay rate due to uncertainty in the antenna locations (i.e.,
fringe search).
On the other hand, since the atmosphere will affect the VLBI phase
stability significantly, we should be careful in selecting the
integration time for VLBI observations, especially in mm/submm wave.
The coherence time is a good indicator of ideal integration time for
the signal detection with VLBI.
This information will be useful for VLBI using single dish telescopes
close to ALMA, namely APEX and ASTE.
Furthermore, VLBI using ALMA (ALMA Phase-Up Project), which is to add
signals from the ALMA antennas in phase in real time to produce a
single VLBI ``station'', or so-called phased array, needs stable
atmospheric phase to obtain high efficiency in adding in phase.
The coherence time is also good to find out the ideal data adding
time length in real time to have high efficiency for phase-up.

\begin{figure}
\plotone{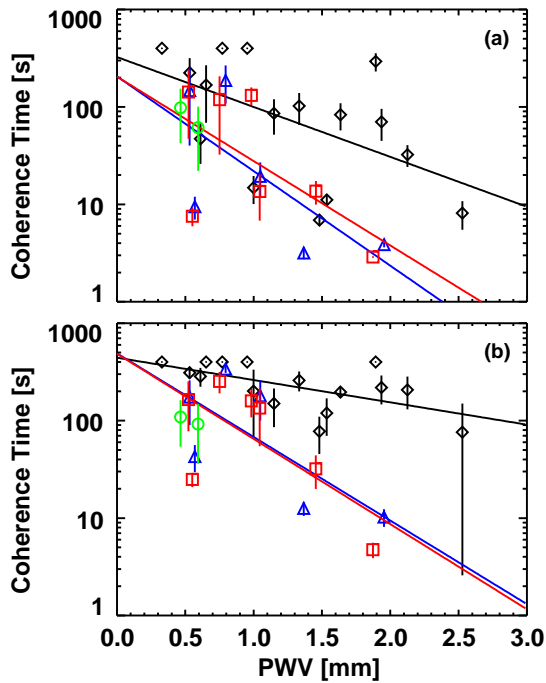}
\caption{Coherence time that lead to a degraded coherence of 0.9
	(i.e., coherence loss of $10\%$) as a function of PWV.
	Each data point in the plot is a median of the coherence time of
	each baseline with its length longer than 1 km in each data set.
	(a) The data before the WVR phase correction, and (b) after.
	Colors for the symbols and the fitted lines are different in
	bands; black, blue, red, and green are for Bands 3, 6, 7, and 8
	data, respectively.
	There is no fitted line for Band 8, since there are not enough
	data points.
	Symbols are the same as in Fig.~\ref{fig-slope}.
\label{fig-coh}}
\end{figure}

Since the phase fluctuation depends on the baseline length, but the
slope will be significantly shallower at baseline lengths longer than
1 km (see Sect.~\ref{sect-ssf}), we take a median value of the
coherence time derived from the data at baseline lengths longer than
1 km.
Here, we calculate the coherence time that degrades the coherence to
0.9 (i.e., coherence loss of 10\%).
The calculation results are shown in Fig.~\ref{fig-coh} as a function
of PWV.
Each calculation stops at the time range of 400 s, since ALMA will
not calibrate phase on such long timescales.
So the data points located at 400 s means that these did not reach
the coherence loss of 10\% even after 400 s.
It is obvious that the data after the WVR phase correction have
longer coherence time than that without.
The average of the coherence time after the WVR phase correction is
184 s, about twice longer than that without (101 s).
This indicates that the WVR phase correction improves the coherence
time by about twice or more (since our calculation stops at 400 s,
the coherence time is underestimated, especially for the WVR phase
corrected data with the time longer than 400 s).
The WVR phase correction is therefore useful for extending
integration or calibration time interval.

In terms of overall PWV dependency of coherence time, there is a weak
trend for lower PWV to give rise to longer coherence time.
This can be understood as more water vapor in the atmosphere, causing
more phase fluctuation and therefore larger coherence loss.
However, the scatter is very large, more than an order of magnitude
of coherence time at a certain PWV, suggesting that phase
fluctuation is not a simple function of the amount of water vapor in
the atmosphere.
This result is the same as Sect.~\ref{sect-ssf-phase}, since we are
showing the same data with different relationship.

The trend appears to depend on frequency band;
Band 3 has larger scatter with less dependence on PWV, but Bands 6
and 7 have smaller scatter with tight dependence on PWV (see the
fitted lines in Fig.~\ref{fig-coh}).
This can be explained as follows; observations at Band 3 can be
carried out under a wider range of PWV, but for Bands 6 and 7, the
observation conditions are limited to better weather and the phase
fluctuation quickly gets worse as weather conditions get worse.

\subsection{Fast Switching Simulation}
\label{sect-fss}

The fast switching phase correction method is to switch between the
science target source and a nearby calibrator quickly (faster than
a time scale on which the phases of both the target and the
calibrator differ from each other),
and calibrate the phase fluctuation \citep{car99,mor00,asa14,asa16}.
The importance of this method is now better understood than it was in
the past.
This is because no matter what causes the phase fluctuation, this
method can improve the phase stability to cancel out not only the
atmospheric phase fluctuations, but also the instrumental phase
errors, such as a frequency standard, which the WVR phase correction
cannot remove.
Compared with the WVR phase correction method, which takes out
shorter time scale phase fluctuation (order of $\sim1$ second to
$\sim1$ minute scale), the fast switching phase correction method
takes out longer time scale phase fluctuation (tens of second or
longer).
Combining those two methods, it is expected to take out a
significant amount of phase fluctuation, and improve the data
quality significantly.

Using the single source stare data taken in 2014, it is possible to
simulate the fast switching phase calibration by separating the data
into ``target'', ``calibrator'', and ``antenna slew time''
components, and making an image for the ``target'' (calibrated using
the ``calibrator'' data).
Comparing the peak flux of the calibrated ``target'' image with that
of the self-calibrated image, namely calculating the coherence of the
data, it is possible to estimate the effect of the fast switching
phase correction method quantitatively by changing the cycle time
(i.e., time for ``calibrator'' $\rightarrow$ ``target'' $\rightarrow$
 ``calibrator,'' including the antenna slew time between the
 calibrator and the target).
This simulation is an idealized case, since the ``target'' and the
``calibrator'' do have the same atmospheric phase fluctuation in this
simulation, which is not true for the real case; it depends on the
separation between the ``target'' and the ``calibrator''
\citep{asa96,asa98}.
This simulation is nonetheless useful for testing 
the effect of cycle time.

\begin{deluxetable}{cccc}
\tablecaption{Time distribution of the cycle time for the fast
	switching simulation.
	\label{tab-fss-ct}}
\tablehead{
	\colhead{Cycle Time}
		& \colhead{Target}
		& \colhead{Calibrator}
		& \colhead{Antenna Slew} \\
	\colhead{[s]}
		& \colhead{[s]}
		& \colhead{[s]}
		& \colhead{[s]}
	}
\startdata
28 & 15 &  5 & 4 \\
48 & 30 & 10 & 4 \\
68 & 45 & 15 & 4
\enddata
\end{deluxetable}

In this paper, we simulated three cycle times: 28 s, 48 s, and 68 s
(see Table~\ref{tab-fss-ct}).
In all cases, we used an integration time for the target of three
times the integration time for the calibrator and a slew time of 4 s,
which are reasonable for the actual observational setup.
Note that the simulated observations always start from and end with
the phase calibrator scan.


We then calibrated the ``target'' data with the ``calibrator'' data,
and imaged the ``target'' using the CASA software and a standard
interferometric imaging procedure.
We also produced the self-calibrated image for the ``target'', and we
used this image as a reference to compare
the peak flux of the simulated image.
The coherence of the fast switching method is estimated by
calculating the ratio of the peak flux densities between the fast
switching simulation image and the self-calibrated one.

\begin{figure}[t]
\plotone{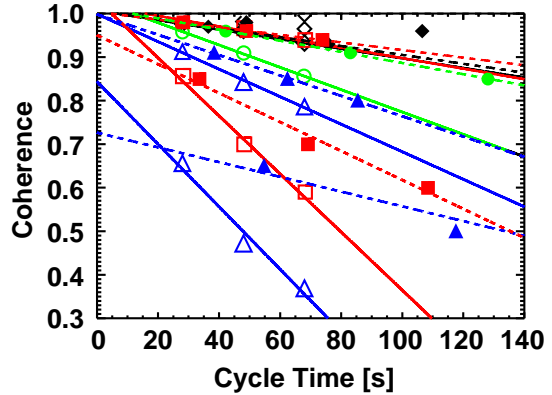}
\caption{Plot of coherence as a function of fast switching cycle time
	for the simulated fast switching method using the single source
	stare data (open symbols).
	Solid lines show the linear fitting results of each data with
	different cycle time.
	We also overplotted the coherence time calculation results using
	the same single source stare data as the fast switching
	simulation data, but for the data points with the baseline
	lengths longer than 1 km (filled symbols).
	Dashed lines are the linear fitting results of each data with
	different coherence time.
	Symbols are the same as in Fig.~\ref{fig-slope}.
	Colors for the symbols and the fitted lines are the same as in
	Fig.~\ref{fig-coh}.
\label{fig-ct}}
\end{figure}

The results of the simulation --- coherence as a function of cycle
time for various frequency bands (Bands 3, 6, 7, and 8) are shown in
Fig.~\ref{fig-ct}.
We also overplotted linear fitting results of coherence as a function
of cycle time as solid lines.
It is obvious that longer cycle time lead to lower coherence.
In addition, the decrease of coherence as a function of cycle time
appears almost linear.
This indicates that fast switching with shorter cycle times could
lead to higher quality images.
This is true irrespective of frequency band and weather conditions.
On the other hand, since we did not consider the separation between
the ``target'' and the ``calibrator'' as mentioned above, this linear
increase of the coherence as a shorter cycle time will be non-linear
or even stop at some point in the case of a real observation, due to
the different atmospheric conditions toward these two sources.
The point at which this happens may highly depends on the separation
between the ``target'' and the ``calibrator,'' and the actual
observational testing of the fast switching phase correction method
using ALMA is currently ongoing \citep{asa14,asa16}.

We also overplot in Fig.~\ref{fig-ct} (filled symbols) coherence as a
function of coherence time, which is calculated as in the previous
subsection (Sect.~\ref{sect-coh}), using the same data sets as the
fast switching simulation.
Here, we set coherence to be the same as that of the fast switching
simulation, and estimated coherence time.
This will be useful once we know the difference between the coherence
time and the simulated fast switching results, since the coherence
time is easily calculated from the quick-look data (i.e., data taken
for a short time, a minute or two, to check the quality of
observational conditions), and it is possible to apply to the actual
observations.
The relation between coherence and coherence time is linear, and the
fitting results are plotted in dashed lines.
Comparison between the results of the fast switching calculations
(open symbols) and that of the coherence time calculations (filled
symbols) shows that the coherence time is almost always about 40-50\%
longer than in the fast switching simulation.
This means that if we estimate the coherence time from the quick-look
data, then the recommended cycle time of an observation with fast
switching will be about 40-50\% shorter.

\section{Summary}
\label{sect-sum}

Using single source stare data taken during the ALMA long baseline
campaigns carried out over the past 6 years, with baseline lengths of
up to $\sim15$ km, we derived various atmospheric phase
characteristics that will be useful for the ALMA long baseline (and
high frequency) observations.
The summary of our study is as follows:
\begin{itemize}
\item The 183 GHz WVR phase correction method works well for reducing
	the phase fluctuation at long baselines, especially for weather
	conditions with precipitable water vapor (PWV) larger than 1 mm.
\item The WVR phase correction lengthens the coherence time by about
	a factor of two or more, indicating that the WVR phase correction
	is useful for lengthening the integration time and/or calibration
	time interval.
\item The WVR phase correction, however, could not take out all the
	phase fluctuation, suggesting that there are other reasons that
	cause phase fluctuation at millimeter and submillimeter
	wavelengths.
	Combining other phase correction methods in addition to the WVR
	phase correction is important, such as the fast switching phase
	correction method.
\item Indeed, our simulation of the fast switching phase correction
	method shows an improvement in the coherence of the data,
	especially with shorter cycle time than a few minutes.
\item Most of the spatial structure functions (SSFs) of the data show
	one turn-over around a baseline length of 1 km.
	This result suggests that the scale height of the turbulent
	constituent is around 1 km, consistent with the distribution of
	water vapor.
	Combined with the success of the WVR phase correction, it is
	obvious that the main turbulent constituent is water vapor in the
	atmosphere.
\item The fitted slopes indicate that most of the phase fluctuation
	at baseline lengths shorter than 1 km shows the intermediate
	value of theoretical 3-D and 2-D Kolmogorov turbulence (i.e.,
	around 0.6), and that at baseline lengths longer than 1 km
	displays the intermediate value of theoretical 2-D Kolmogorov
	turbulence and no correlation (i.e., $0.2-0.3$).
\item There are a few cases that do not show any turn-over.
	Such SSFs are only obtained under very low PWV conditions, when
	the improvement factor of the WVR phase correction is not high,
	or after the WVR phase correction.
	This suggests that the main turbulent constituent in these cases
	is not water vapor, and that the phase fluctuation caused by this
	constituent is inherently smaller than that caused by water
	vapor.
	The scale height of this constituent is higher than 10 km,
	suggesting that water ice or a dry component (N$_{2}$ or O$_{2}$)
	can be the cause of the phase fluctuation.
\item Excess path length fluctuation at a baseline length of 10 km
	is large, $\sim200~\mu$m even at PWV less than 1 mm (no large
	difference for the data before and after the WVR phase
	correction), and increases as a function of PWV.
	This value is significant for the high frequency observations,
	and strongly suggest to use other phase correction method, such
	as the fast switching and/or band-to-band phase correction
	methods.
\end{itemize}

\acknowledgements

This paper makes use of the following ALMA data:
ADS/JAO.ALMA\#0000.0.00341.CSV.
ALMA is a partnership of ESO (representing its member states), NSF
(USA) and NINS (Japan), together with NRC (Canada) and NSC and ASIAA
(Taiwan), and KASI (Republic of Korea), in cooperation with the
Republic of Chile.
The Joint ALMA Observatory is operated by ESO, AUI/NRAO and NAOJ.

We express our gratitude to all the ALMA members for the support of
the long baseline campaigns.
SM and YA thank the Joint ALMA Observatory (JAO) for supporting our
visit to ALMA as Expert Visitors in 2013 and 2014.
SM and YA also expresses their appreciations to the National
Astronomical Observatory of Japan (NAOJ) for their supports for their
stays in Chile in 2012 -- 2014.
SM is supported by the National Science Council (NSC) and the
Ministry of Science and Technology (MoST) of Taiwan,
NSC 100-2112-M-001-006-MY3 and MoST 103-2112-M-001-032-MY3.
LM and RT are part of the Dutch ALMA ARC node, Allegro, which is
funded by the Netherlands Organisation for Scientific Research (NWO).

We deeply regret the loss of our colleague, Koh-Ichiro Morita, who
lost his life on May 7th, 2012, at Santiago, Chile.
He significantly helped not only this work but also for various works
in the ALMA project, and without him, we could not achieve these
results.
Furthermore, he taught S.M.\ the basic and application of
interferometry, phase correction methods, and site testings since
S.M.\ was a graduate student at the Nobeyama Radio Observatory.
Without his guidance, S.M.\ could not be at the current position.
Morita-san, we really miss you...

\appendix
\section{Data List}
\label{sect-app-data}

Here, we present the list of the data used in this study.
Data for the longest baseline lengths of 2 km, 3 km, and $10-15$ km
are shown in Tables~\ref{tab-app-2km}, \ref{tab-app-3km}, and
\ref{tab-app-10km}, respectively.

\begin{deluxetable}{lccllclc}
\tablecaption{List of the 2 km baseline data used in this study.
	\label{tab-app-2km}}
\tablehead{
	\colhead{Date}
		& \colhead{Time}
		& \colhead{Duration [min]}
		& \colhead{ExecBlock}
		& \colhead{PWV}
		& \colhead{Band}
		& \colhead{Source}
		& \colhead{No.\ of BLs} \\
		& \colhead{[UT]}
		& \colhead{[min]}
		&
		& \colhead{[mm]}
		&
		&
		& \colhead{with > 1 km}
	}
\startdata
2012/05/02 & 11:02 & 20 & uid\_\_\_A002\_X3f4b85\_X392 & 1.54 & 6 & 1924-292 & 13 \\
           & 11:28 & 20 & uid\_\_\_A002\_X3f4b85\_X399 & 1.65 & 6 & 1924-292 & 13 \\
2012/05/10 & 04:22 & 20 & uid\_\_\_A002\_X3ffc69\_X71  & 0.98 & 6 & 3C279    & 14 \\
           & 10:45 & 20 & uid\_\_\_A002\_X401024\_Xca  & 0.86 & 6 & 2258-279 & 13 \\
           & 22:10 & 20 & uid\_\_\_A002\_X401c5a\_X13e & 1.96 & 3 & 0522-364 & 16 \\
2012/05/11 & 09:13 & 20 & uid\_\_\_A002\_X403046\_X3d  & 1.78 & 3 & 1924-292 & 14 \\
2012/05/12 & 00:00 & 20 & uid\_\_\_A002\_X404191\_X26  & 1.91 & 3 & 3C279    & 16 \\
           & 12:13 & 20 & uid\_\_\_A002\_X4056cb\_X3ac & 1.14 & 3 & 3C454.3  & 13 \\
           & 12:36 & 20 & uid\_\_\_A002\_X4056cb\_X3b3 & 1.30 & 7 & 3C454.3  & 13 \\
2012/05/15 & 23:23 & 20 & uid\_\_\_A002\_X40b03b\_X2c  & 1.69 & 3 & 3C279    & 20 \\
2012/05/16 & 01:29 & 20 & uid\_\_\_A002\_X40b03b\_X36c & 1.56 & 3 & 3C279    & 19 \\
           & 02:39 & 20 & uid\_\_\_A002\_X40b03b\_X465 & 1.46 & 3 & 3C279    & 18 \\
           & 12:17 & 20 & uid\_\_\_A002\_X40c17a\_X120 & 0.69 & 3 & 3C454.3  & 18 \\
2012/05/26 & 05:53 & 20 & uid\_\_\_A002\_X412ad5\_X5dd & 1.97 & 3 & 1924-292 & 15
\enddata
\end{deluxetable}

\begin{deluxetable}{lccllclc}
\tablecaption{List of the 3 km baseline data used in this study.
	\label{tab-app-3km}}
\tablehead{
	\colhead{Date}
		& \colhead{Time}
		& \colhead{Duration [min]}
		& \colhead{ExecBlock}
		& \colhead{PWV}
		& \colhead{Band}
		& \colhead{Source}
		& \colhead{No.\ of BLs} \\
		& \colhead{[UT]}
		& \colhead{[min]}
		&
		& \colhead{[mm]}
		&
		&
		& \colhead{with > 1 km}
	}
\startdata
2013/06/03 & 22:10 & 10 & uid\_\_\_A002\_X65a644\_Xd    & 0.68 & 3 & 1058+015 & 46 \\
2013/06/06 & 05:59 & 10 & uid\_\_\_A002\_X660b54\_Xae2  & 0.54 & 3 & NRAO530  & 49 \\
           & 09:00 & 10 & uid\_\_\_A002\_X660b54\_X16de & 0.55 & 3 & 1924-292 & 47 \\
           & 11:12 & 10 & uid\_\_\_A002\_X660b54\_X1caf & 0.55 & 3 & 2348-165 & 43 \\
           & 11:33 & 10 & uid\_\_\_A002\_X660b54\_X1d7e & 0.54 & 6 & 2348-165 & 43 \\
           & 11:51 & 10 & uid\_\_\_A002\_X660b54\_X1f3c & 0.57 & 7 & 2348-165 & 43 \\
2013/06/07 & 05:24 & 10 & uid\_\_\_A002\_X6643c8\_X3e6  & 1.46 & 3 & 1924-292 & 53 \\
           & 05:43 & 10 & uid\_\_\_A002\_X6643c8\_X425  & 1.42 & 6 & 1924-292 & 53 \\
           & 06:00 & 10 & uid\_\_\_A002\_X6643c8\_X47d  & 1.39 & 7 & 1924-292 & 53
\enddata
\end{deluxetable}

\begin{deluxetable}{lccllclc}
\tablecaption{List of the $10-15$ km baseline data used in this study.
	\label{tab-app-10km}}
\tablehead{
	\colhead{Date}
		& \colhead{Time}
		& \colhead{Duration [min]}
		& \colhead{ExecBlock}
		& \colhead{PWV}
		& \colhead{Band}
		& \colhead{Source}
		& \colhead{No.\ of BLs} \\
		& \colhead{[UT]}
		& \colhead{[min]}
		&
		& \colhead{[mm]}
		&
		&
		& \colhead{with > 1 km}
	}
\startdata
2014/09/09 & 01:41 & 30 & uid\_\_\_A002\_X8b5a69\_Xd3   & 1.64 & 3 & 2253+1608 & 260 \\
           & 02:17 & 30 & uid\_\_\_A002\_X8b5a69\_X233  & 1.45 & 7 & 2253+1608 & 260 \\
2014/09/12 & 07:26 & 30 & uid\_\_\_A002\_X8ba346\_X9bb  & 1.89 & 3 & 0428-3756 & 222 \\
2014/09/13 & 06:32 & 40 & uid\_\_\_A002\_X8bc8aa\_X962  & 2.53 & 3 & 0522-364  & 181 \\
2014/09/14 & 09:25 & 40 & uid\_\_\_A002\_X8bd3e8\_Xa59  & 0.61 & 3 & 0522-364  & 259 \\
           & 10:16 & 40 & uid\_\_\_A002\_X8bd3e8\_Xac6  & 0.57 & 6 & 0522-364  & 259 \\
	       & 11:22 & 40 & uid\_\_\_A002\_X8bd3e8\_Xc55  & 0.55 & 7 & 0522-364  & 245 \\
2014/09/15 & 08:02 & 40 & uid\_\_\_A002\_X8beb08\_Xe43  & 0.95 & 3 & 0522-364  & 262 \\
           & 08:50 & 40 & uid\_\_\_A002\_X8beb08\_Xea2  & 0.98 & 7 & 0522-364  & 262 \\
2014/09/16 & 07:25 & 40 & uid\_\_\_A002\_X8c0e65\_Xc72  & 0.77 & 3 & 0522-364  & 278 \\
           & 08:11 & 40 & uid\_\_\_A002\_X8c0e65\_Xe16  & 0.80 & 6 & 0522-364  & 265 \\
           & 10:36 & 40 & uid\_\_\_A002\_X8c0e65\_X1151 & 0.75 & 7 & 0522-364  & 276 \\
2014/09/23 & 01:29 & 40 & uid\_\_\_A002\_X8d054c\_X276  & 2.13 & 3 & 3C454.3   & 447 \\
           & 05:00 & 40 & uid\_\_\_A002\_X8d054c\_X99b  & 1.95 & 6 & 3C454.3   & 360 \\
           & 05:45 & 40 & uid\_\_\_A002\_X8d054c\_Xaf4  & 1.87 & 7 & 3C454.3   & 378 \\
2014/09/26 & 23:51 & 30 & uid\_\_\_A002\_X8d8bc2\_Xe5   & 1.15 & 3 & 1924-292  & 378 \\
2014/09/27 & 00:28 & 30 & uid\_\_\_A002\_X8d8bc2\_X125  & 1.05 & 6 & 1924-292  & 378 \\
           & 01:08 & 30 & uid\_\_\_A002\_X8d8bc2\_X165  & 1.04 & 7 & 1924-292  & 378 \\
           & 19:18 & 30 & uid\_\_\_A002\_X8da59e\_X5    & 1.48 & 3 & 1924-292  & 440 \\
           & 19:53 & 30 & uid\_\_\_A002\_X8da59e\_X45   & 1.37 & 6 & 1924-292  & 440 \\
           & 20:34 & 30 & uid\_\_\_A002\_X8da59e\_X85   & 1.54 & 3 & 1924-292  & 440 \\
2014/09/28 & 16:21 & 30 & uid\_\_\_A002\_X8dd128\_X2b   & 1.00 & 3 & 3C279     & 191 \\
2014/09/29 & 06:32 & 30 & uid\_\_\_A002\_X8dd2a4\_Xae7  & 0.59 & 8 & 0522-364  & 280 \\
2014/10/01 & 05:27 & 30 & uid\_\_\_A002\_X8e1004\_Xec4  & 0.53 & 3 & 0522-364  & 400 \\
           & 06:06 & 30 & uid\_\_\_A002\_X8e1004\_Xf04  & 0.53 & 6 & 0522-364  & 370 \\
           & 06:43 & 30 & uid\_\_\_A002\_X8e1004\_Xf44  & 0.52 & 7 & 0522-364  & 400 \\
           & 08:02 & 90 & uid\_\_\_A002\_X8e1004\_Xfe5  & 0.46 & 8 & 0522-3627 & 360 \\
2014/10/05 & 09:37 & 30 & uid\_\_\_A002\_X8ec7bb\_X1123 & 0.33 & 3 & 0522-364  & 286 \\
2014/10/29 & 23:06 & 30 & uid\_\_\_A002\_X91bdc6\_X36   & 1.94 & 3 & 3C454.3   & 410 \\
2014/10/30 & 02:43 & 30 & uid\_\_\_A002\_X91cc20\_X8f   & 0.65 & 3 & 3C454.3   & 473 \\
2014/11/01 & 19:04 & 30 & uid\_\_\_A002\_X920302\_X1d6d & 1.33 & 3 & 1924-292  & 410
\enddata
\end{deluxetable}

\end{document}